\documentclass[a4paper,12pt]{article}

\usepackage[utf8]{inputenc}
\usepackage[english,russian]{babel}
\usepackage{amssymb,amsmath}
\usepackage[pdftex]{graphicx}
\usepackage{hyperref}
\usepackage{color}
\normalsize

\begin{document}

\thispagestyle{empty}
\begin{flushright}
\begin{tabular}{l}
\textbf{JOINT}\\
\textbf{INSTITUTE OF}\\
\textbf{NUCLEAR}\\
\textbf{RESEARCH}\\
\hspace{2.5cm}\textbf{DUBNA}	
\end{tabular}
\end{flushright}
\begin{flushleft}
	\includegraphics[scale=0.5]{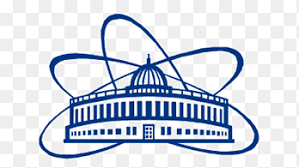}
\end{flushleft}
\hrule
\vspace{1.5cm}
\begin{flushright}
	\textbf{P2 -- 11308}
\end{flushright}
\vspace{2cm}
{\large
\hspace{2cm}\textbf{Gerdt V.~P., Karimkhodzhaev A., Faustov R.~N.}}
\vspace{2.5cm}

\hspace{1.6cm}\texttt{\Large HADRONIC VACUUM POLARIZATION}

\hspace{1.6cm}\texttt{\Large AND TEST OF QUANTUM ELECTRODYNAMICS}

\hspace{1.6cm}\texttt{\Large AT LOW ENERGIES}

\vspace{3cm}
\begin{flushright}
	\textbf{\Huge 1978}
\end{flushright}
\hrule
\newpage
\thispagestyle{empty}
{\renewcommand{\thefootnote}{\fnsymbol{footnote}}
.
\vspace{8cm}
\begin{flushright}
	\textbf{P2 -- 11308}
\end{flushright}
\vspace{2cm}
{\large
\hspace{2cm}\textbf{V.P.GERDT}, \textbf{A.KARIMKHODZHAEV\footnote[1]{\textbf{\large INP AS UzSSR, Tashkent}}}, \textbf{R.N.FAUSTOV}}
\vspace{2.5cm}

\hspace{1.6cm}\texttt{\Large HADRONIC VACUUM POLARIZATION}

\hspace{1.6cm}\texttt{\Large AND TEST OF QUANTUM ELECTRODYNAMICS}

\hspace{1.6cm}\texttt{\Large AT LOW ENERGIES}

\vspace{3cm}

\hspace{1.6cm}\textit{\large \textbf{Sent to the collection ''Problems of the theory of gravitation}} 

\hspace{1.6cm}\textit{\large \textbf{and elementary particles ''}}}
\clearpage

\newpage

\begin{center}
\begin{tabular}{|p{16cm}|}

\\
\hline
Gerdt V.~P., Karimkhodzhaev A., Faustov R.~N.\qquad P2 - 11308\\
\\
\\
Hadronic Vacuum Polarization and Test of Quantum Electrodynamics at
Low Energies\\
A hadronic vacuum polarization correction to the photon propagator
is found by using the Dubnicka-Meshcheryakov parameterization of the
pion electromagnetic form factor and new experimental data on the
$e^+e^-$ hadrons annihilation cross section. Then, the contribution
from the hadronic vacuum polarization to the muon anomalous magnetic
moment and the Lamb shift in muonic atoms are calculated.\\
The investigation has been performed at the Laboratory of
Theoretical Physics, JINR\\
\\
\\
Preprint of the Joint Institute for Nuclear Research. Dubna 1978\\
\\
\hline
\end{tabular}

\end{center}

\copyright 1978 Joint Institute for Nuclear Research. Dubna 1978\\
\newpage

\section*{1. INTRODUCTION}

The test of quantum electrodynamics (QED)  at  the small distance is one of the important problem in the elementary particle physics. Many experemental and theoretical difficulties emerge in the region of small space interval . One of theoretical difficulties, for example, is the necessary of considering contributions of higher order in coupling constant   $\alpha=\frac{e^2}{4\pi}=\frac{1}{137}$.  In order to evaluate the deviations  from predictions  of the
"pure" QED , it is negessary to know corrections to observables  due to strong
and weak interactions . Particularly , it is necessary to consider the effect of the
structure  of hadrons. But this encounters serve difficulties , since we do not
have any appropriate theory of strong interactions .The applicability of QED
can be analized by two means : the first is to do experiments with large momentum transfer of high energy particles  (experiments with colliding $e^+e^-$ -beams,
collision of leptons with nucleons ).The second is to measure low energy
observables such as the fine and superfine structures of atomic levels and
anomalous magnetic moments of electron and muon. The influence of strong
interections on the qbserved quantities is manifested, in addition to taking into account the structure of hadrons, also through the hadronic vacuum  polarization (HVP)
    In this note we will find the correction by HVP to the photon propagator, and with its help we will calculate the contribution HVP to the anomalous magnetic moment of the muon and Lamb shift of muonic atoms.

\section*{2. THE CONTRIBUTION OF HADRONIC VACUUM POLARIZATION INTO PHOTON GREEN FUNCTION}

The photon Green's function in the transverse gauge has the form\cite{1}
\begin{equation} \label{1} 
D^{\mu\nu}(q^2)=-\left(g^{\mu\nu}-\frac{q^{\mu} q^{\nu}}{q^2}\right)D(q^2).
\end{equation}

Where the invariant function $D(q^2)$ is expressed by using the HVP operator$\Pi^h(q^2)$ in the  $e^2$- approximation as follows (Fig .1):
\begin{equation}\label{2} D(q^2)=\frac{1}{q^2}[1-\Pi^h(q^2)].
\end{equation}

%
%
\begin{figure}[h!]
\begin{center}
\includegraphics[width=0.9\linewidth]{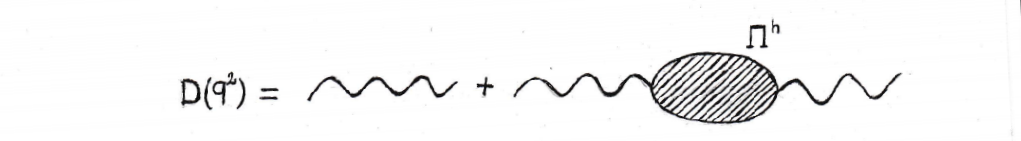}\\
{Fig.1}
\end{center}
\end{figure}
%
%

   We wraite the invariat function  $\Pi^h(q^2)$ in the Callan-Lehmann \cite{1} representation with  one supraction
\begin{equation}\label{3}
\Pi^h(q^2)=e^2q^2\int^\infty_{4m^2_\pi}\frac{\rho^h(s)ds}{s(s-q^2-i0)},
\end{equation}

where the hadronic spectral function $\rho^h$ is connected with the total $e^+e^-$ annihilation cross section into hadrons by the folloving relation (Fig.2)

\begin{equation}\label{4} 
\sigma^h(q^2)=\frac{16\pi^3\alpha^2}{q^2}\rho^h(q^2)
\end{equation}

%
%
\begin{figure}[h!]
\begin{center}
\includegraphics[width=0.9\linewidth]{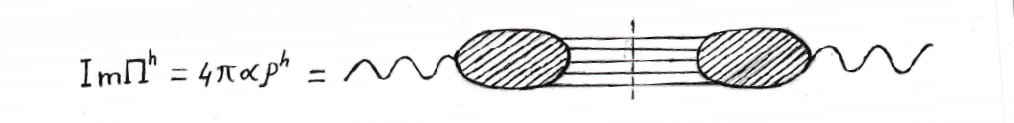}\\
{Fig.2}
\end{center}
\end{figure}
%
%


 As usual ,it is convenient to introduce the ratio of cross section
\begin{equation}
\label{5} 
R=\frac{\sigma_h(e^+e^- \rightarrow \{hadrons\})}
{\sigma_{\mu\mu}(e^+e^-\rightarrow \mu^+\mu^-)},
\end{equation}
where
$$
\sigma_{\mu\mu}=\frac{4\pi}{3}\frac{\alpha^2}{s}.
$$

 Then for the spectral function $\rho^h$ we have
$$
\rho^h(q^2)=\frac{1}{12\pi^2}\frac{\sigma_h}{\sigma_{\mu\mu}}=\frac{1}{12\pi^2}R
$$
As a result, the photon propogator $D(q^2)$  taking into account  the  HVP corection in the  $e^2$-approximation, takes the following form  /2/\footnote{Sum rules for $\Pi(q^2)$  in theory with asymptotic freedom were concidered in \cite{21}}:

\begin{equation}\label{6} D(q^2)=\frac{1}{q^2+i0}\left[1-\frac{\alpha}{3\pi}q^2
\int^\infty_{4m^2_\pi}\frac{R(s)ds}{s(s-q^2-i0)}\right]
\end{equation}

Since the integral in equation {(6)} converges guickly, it is clear that the essential
contribution to the HVP comes from the region near the threshold, i.e. the two
pion intermediate  state. Our  job is to find the parameterization
  of $R$ by using the improved description of experimental data in the region of two pion creation \cite{3} and by taking into account of new structures  \cite{4} in the   $\sigma_h(e^+e^-\rightarrow {hadrons})$  cross section.\\

Cross section  $\sigma_h(e^+e^-\rightarrow {hadrons})$ can be represented as the  sum of several parts :
\begin{equation}\label{7} 
\sigma_h=\sigma_{background}+\sigma_{resonans}+\sigma_{heavy lepton}
\end{equation}

First term- $\sigma_{background}$ determined by fitting experimental \cite{3}-\cite{5} points after subtracting the resonance contribution. The second term $ \sigma_{resonans}$  is expressed by generalized vector dominance \cite{6} model, by the following Breit - Wigner representation: 

\begin{equation}\label{8}
\sigma_{{resonance}}=\frac{12\pi}{s}\sum\limits_i
\frac{m^2_i\Gamma_i\Gamma^\ell_i}{(s-m^2_i)+m^2_i\Gamma^2_i},
\end{equation}

where $i$ means both ''old'' $(\rho,\omega,\phi$, and ''new''
$(J/\psi,\psi',\psi'',\ldots)$  vector mesons having the quantum numbers of photon; ; $m_i,\Gamma_i,\Gamma^\ell_i$ -  respectively mass, total and leptonic width of vector mesons. Third term $\sigma_{heavy lepton}$ - heavy lepton contribution to the total cross section from decays into hadrons of a recently discovered  \cite{7}  $\tau$ with mass  $m=1,8$ GeV, wich for  $s\gg m^2_\tau$ has the form 

\begin{equation}\label{9}
\sigma_{\tau\tau}\cong\sigma_{\mu\mu}=\frac{4\pi}{3}\frac{\alpha^2}{s},
\end{equation}
and correspondingly,

\begin{equation}\label{10} 
R=\frac{\sigma_{\tau\tau}}{\sigma_{\mu\mu}}\cong{1}.
\end{equation}

At the treeshold of heavy lepton production for  $R_\tau$ the follow value is obtained \cite{7}

\begin{equation}\label{11} R_\tau=0,89\pm 0,29\pm 0,27,
\end{equation}
where the first error is statistical  and second is sistematic. Since the main contribution    
to the integral {(6)} comes from $\rho$ -meson and the two pion state , it is natural use more precise  expression in this area for $\sigma(e^+e^-\rightarrow\pi^+\pi^-)$, than the formula {(8)}. The cross section  $\sigma(e^+e^-\rightarrow\pi^+\pi^-)$
is expressed by pion electromagnetic form factor  $F_{\pi}(s)$ \cite{3,8}
\begin{equation}\label{12}
\sigma(e^+e^-\rightarrow\pi^+\pi^-)=\frac{8\pi}{3}\frac{\alpha^2q^2}{s^{5/2}}|F_{\pi}(s)|^2,
\end{equation}
where
\begin{equation}\label{13} q=(\frac{s}{4}-m^2_{\pi})^{1/2}.
\end{equation}

There are varios parameterizations  \cite{9,10}  for form factor  $F_{\pi}(s)$. 
In calculations we used the  Dubnicka -Meshcheryakov parameterization \cite{10}, which destribes experimental data in the wide range of energy $-2$ GeV$^2\leq s\leq 4,4$ GeV$^2$, possessing an asymptotic behavior $1/s$, which is consistant with the results of the rules quark counting  \cite{11}. In this specified parameterization of the form factor, $F_{\pi}(s)$ is as follows:

\begin{equation}\label{14}
F_{\pi}(s)=P_1(s)\frac{(q-q_1)}{(q+q_2)(q+q_3)(q+q_4)}\frac{(i+q_2)(i+q_3)(i+q_4)}{(i-q_1)},
\end{equation}

where \footnote{The numerical values of the parameters  $A,\,q_i$, correspond to unit mass of a pion $(m_{\pi}=1)$.}
\begin{eqnarray}
	P_1&=&1+A\cdot s,\nonumber\\
	A&=&0,0027\pm 0,0003,\nonumber\\
	q_1&=&-i0,960504,\nonumber\\
	q_2&=&-2,565913+i0,289811,\nonumber\\
	q_3&=&i1,048006,\nonumber\\
	q_4&=&2,565913+i0,289811.\nonumber
\end{eqnarray}

Interference  $\rho$ и $\omega$ up to the production threshold 
$\omega\pi-0,92$ GeV  we took into account \cite{3} by adding to $F_{\pi}(s)$
 the expression
 
\begin{equation}\label{15}
F^{\omega}_{\pi}=\frac{i0,014}{1-\frac{s}{m^2_{\omega}}-i\frac{\Gamma^{\omega}}{m_\omega}}.
\end{equation}

For the rest of resonances we used representation Breit - Wigner \cite{8}. 
To determine the contribution from the  $\sigma_{background}$    to $R$, we have taking into account the fact that the experimental  value of  $R$ descreases slowly up to  $\sqrt{s}=3,0$  GeV  and increases in the ragion $3,0  GeV\leq\sqrt{s}\leq 5,0  GeV$  due to the production of charm particles and starting  from $\sqrt{s}=5,0$ GeV  becomes approximately constant and equal to $4,5-5,5$. Fitting of experimental data \cite{3}-\cite{5} according to the formula $R=A\cdot s^n$  by the least squars  method gives:
 \vspace{0,2cm} \\
 \begin{tabular}{ll}
$ R=(5,4\pm1,0)\cdot s^{-(0,5\pm0,1)}\quad$& in the region $1,2\leq\sqrt{s}\leq 3,0$ Gev,
\\
 $R=(0,4\pm0,1)\cdot s^{(0,8\pm0,1)}\quad$& in the region $3,0\leq\sqrt{s}\leq 5,0$ GeV and \\
$R=(5,3\pm0,5)$& in the region $5,0\leq\sqrt{s}\leq 7,4$ Gev.
\end{tabular}

\noindent
For asymptotics $\sqrt{s}>7,4$ GeV  we can use the formula following from asymptotic freedom assumptions \cite{12}
\begin{equation}\label{16}
R_\infty=(3\sum_iQ^2_i)(1+\frac{\delta}{ln(\frac{s}{\Lambda^2})}),
\end{equation}

where $Q_i$-is the quark charge , $\delta=\frac{12}{25}$   for the model with four colored quarks, 
\begin{center}
$\Lambda={(5,4\pm0,6)}$ GeV.
	\end{center}
	Now let's move on to specific applications in which the correction for the HVP to the photon propagator affects.

\section*{3. CONTRIBUTION OF HVP TO ANOMALOUS MAGNETIC MOMENT OF THE MUON }

Recent measurements of anomalous magnetic moment  $-a_\mu$   of the muon give the result  \cite{13}
\[
a_\mu{(exp.)}=(1165922\pm9)\cdot10^{-9}(8ppm).
\]

Theoretical value $a_\mu$, including calculations up to the eighth order in  $\alpha$ \cite{13,14},  equals 
\[a_{\mu}{(QED)}=(1165851,8\pm 2,4)\cdot10^{-9}.\]
The difference between the theoretical value of $a_{\mu}$(QED) from experimental

\begin{equation}\label{17} 
a_\mu{(exp.)}-a_{\mu}{(QED)}=(70,2\pm 11,4)\cdot 10^{-9}
\end{equation}

can be explained by the presence of HVP effects.
можно объяснить наличием эффектов АПВ. The hadronic contribution to  $a_\mu$,
which is manifested by the correction on HVP to the photon propagator  {(6)},
expressed through the well known integral  \cite{14,15} /Fig. 3/
\begin{equation}
\label{18} 
a_\mu\mbox{(эксп.)}=(\frac{m_{\mu}\alpha}{3\pi})^2
\int^{\infty}_{4m^2_\pi}\frac{R(s)K(s)}{s^2}ds,
\end{equation}
\\
%
%
\begin{figure}[h!]
\begin{center}
\includegraphics[width=0.5\linewidth]{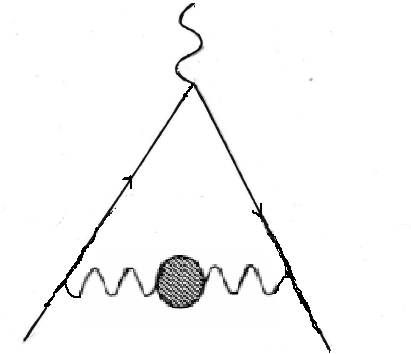}\\
{Fig.3}
\end{center}
\end{figure}
%
%

where $m_\mu$ - mass of $\mu$ -meson. The kernel $K(s)$ has the form \cite{15}
$$
K(s)=\frac{3s}{m^2_{\mu}}\left\{\frac{1}{2}x^2(2-x^2)+
\frac{(1+x^2)(1+x)^2}{x^2}\left[\ln(1+x)-x+\frac{x^2}{2}+
\left(\frac{1+x}{1-x}\right)x^2\ln(x)\right]\right\},
$$
where
$$x=\frac{s}{4m^2_\mu}(1-\sqrt{1-\frac{4m^2_\mu}{s}})^2.$$
Notice, that
\[
K(s\rightarrow\infty )=1
\] and 
\[
K(s\cong m^2_\rho)\cong 0,877.
\]
Calculation of the integral {(18)} using parameterization $R$, given in Section 2 gives the result
\begin{equation}\label{19} 
a_\mu{(hadrons)}=(72,15\pm 5,96)\cdot10^{-9},
\end{equation}
which is in good agreement with earlier   calculations \cite{14,16}  and experiment  (Table.1.)

\begin{table}[h!]
\begin{center}
\begin{tabular}{|l|c|}
	\hline
	Contributions to $a_\mu$(hadrons)&$\times 10^{-9}$\\
	\hline
	$\rho,\omega\rightarrow 2\pi,\sqrt{s}<0,92$&47,3$\pm$0,70\\
	$\omega\rightarrow 3\pi$&5.10$\pm$1,16\\
	$\varphi$&4,10$\pm$0,32\\
	$J/\Psi$(3,095), $\Psi'$(3.684), $\Psi''$(3,772),
	$\Psi$(4,4)&0,70$\pm$0,18\\
	background $1,2\leq\sqrt{s}\leq 3,0$&12,1$\pm$3,10\\
	background $3,0<\sqrt{s}\leq 5,0(a_\mu(\tau)=0,22\pm 0,07)$&0,70$\pm$0,32\\
	background $5,0<\sqrt{s}\leq7,4(a_\mu(\tau)=0,13\pm 0,04)$&0,50$\pm$0,07\\
	Asymptotics $\sqrt{s}>7,4(a_\mu(\tau)=0,12)$&\\
	(asymptotic freedom)&0,60$\pm$0,05\\
	\hline
	Total &72,15$\pm$5,96\\
	\hline
\end{tabular}
\end{center}
\begin{center}
Table.1
\end{center}
\end{table}

Note that the contribution to the  $a_\mu$ pion form factor for the parameterization of Dubnichka-Meshcheryakov slightly higher than for the Gunnaris-Sakurai parameterizations \cite{9,16}, which is equal to $46,5\cdot 10^{-9}$. Contribution from $\sigma_{background}$, in the region  $1,2<\sqrt{s}< 3,0$ GeV is also slightly higher than others authors, but the difference is within the measurement  error. Generally speaking, more than half of the error is in the value $a_\mu$
based on errors of experimental data in the region up to   $\sqrt{s}= 3,0$ GeV,
therefore it would be desirable to careful measurement of the cross section
$\sigma_h$ up to $\sqrt{s}= 3,0$ Gev.\\

The contribution of heavy lepton to  $a_\mu${(hadrons)} is equal to
\[a_\mu{(heavy lepton)}=(0,47\pm 0,11)\cdot10^{-9}\]
As shown in article \cite{14}, the contribution of higher-order HVP  by $\alpha$
to $a_\mu$ is equal to
\[(-3,5\pm1,4)\cdot10^{-9}.\]
Adding this contribution to the expression \cite{19}, we obtain
\[a_\mu{(hadrons)}=(68,65\pm7,36)\cdot10^{-9},\]
which is in good agreement with the  \cite{17}.

\section*{4.  CONTRIBUTION OF HADRONS TO LAMB SHIFT IN MUONIC ATOMS }

 In the case when  $q^2\ll4m^2_\pi$ the  formula  {(6)} may be    rewritten as
\begin{equation}\label{20}
D(q^2)\cong\frac{1}{q^2}[1-\frac{\alpha}{3\pi}\frac{q^2}{m^2_h}],
\end{equation}
where $m^2_h$ -  is the effective hadron mass defined as
\begin{equation}
\label{21} m^{-2}_h=\int^\infty_{4m^2_h}\frac{R(s)}{s^2}ds.
\end{equation}

Numerical integrations in  {(21)} using the parameterization $R$ given in section 2 lead to the results collected in Table. 2.
\begin{table}[h!]
\begin{center} 	
\begin{tabular}{|l|c|}
	\hline
	Contributions to $a_\mu$(hadrons)&$\times 10^{-9}$\\
	\hline
	$\rho,\omega\rightarrow 2\pi,\sqrt{s}<0,92$&16,5$\pm$0,30\\
	$\omega\rightarrow3\pi$&1,70$\pm$0,40\\
	$\varphi$&1,40$\pm$0,10\\
	$J/\Psi$(3,095), $\Psi'$(3.684), $\Psi''$(3,772),
	$\Psi'''$(4,4)&0,23$\pm$0,05\\
	background $1,2\leq\sqrt{s}\leq 3,0$&4,70$\pm$1,05\\
	background $3,0<\sqrt{s}\leq5,0(m^2_\tau=0,06\pm0,02)$&0,24$\pm$0,10\\
	background $5,0<\sqrt{s}\leq7,4(m^2_\tau=0,04\pm0,01)$&0,20$\pm$0,02\\
	Asimptotics $\sqrt{s}>7,4(m^2_\tau=0,03)$&\\
	(asymptotic freedom)&0,15$\pm$0,01\\
	\hline
	Total &25,12$\pm$2,03\\
	\hline
\end{tabular}
\end{center}
\begin{center}
Table.2
\end{center}
\end{table}\\
Then:
\begin{equation}\label{22} 
m^{-2}_h=\frac{0,25\pm0,02}{m^2_\pi}
\end{equation}

or $m^2_h\approx 4m^2_\pi.$\\
Taking this value of   $m^{-2}_h$ into account, the HVP contribution to the photon
propagator is expressed by the formula:

\begin{equation}\label{23}
D(q^2)\cong\frac{1}{q^2}[1-\frac{\alpha}{3\pi}\frac{q^2}{4m^2_h}]
\end{equation}

For comparison, note that the contribution of only one $\pi$-  meson loop ( Fig. 4 ) with exact vertexes of one order lower is

\begin{equation}\label{24}
D(q^2)\cong\frac{1}{q^2}\left[1-\frac{\alpha}{3\pi}\frac{q^2}{40m^2_\pi}\right],
\end{equation}
e.g.  $m^2_h\simeq40m^2_\pi$.\\
%
%
\begin{figure}[h!]
\begin{center}
\includegraphics[width=0.8\linewidth]{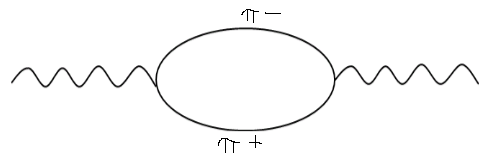}\\
{Fig.4}
\end{center}
\end{figure}
%
%
The potential of interaction with the correction from HVP has the form

\begin{equation}\label{25} 
V(\vec{q}\,^2)=-Ze^2\frac{d(q^2)}{\vec{q}\,^2},\quad
q^2=-\vec{q}\,^2
\end{equation}

where $Z$ - the charge of atomic neclei and   $d(q^2)=q^2D(q^2)$. 
Let us separate the pure Coulomb potential and write

\begin{equation}\label{26} V({\vec{q}}\,^2)=V_C({\vec{q}}\,^2)+\Delta V({\vec{q}}\,^2),
\end{equation}

where $V_C(\vec{q}\,^2)=-\frac{Ze^2}{\vec{q}\,^2}$ - is the Coulomb potential,  $\Delta V(\vec{q}\,^2)$ is HVP  correction  to the potential. 
 
Consequently, it has the form

\begin{equation}\label{27} \Delta
V({\vec{q}}\,^2)=-\frac{Ze^2}{{\vec{q}}\,^2}[d(q^2)-1]=-\frac{4\alpha}{3}\frac{Z\alpha}{m^2_h}.
\end{equation}

     The corresponding shift of muon atom S-level in the first order of
perturbation theory is

\begin{equation}
\label{28} \Delta E_n=\langle\Psi_{C}|\Delta V(\overrightarrow{r})|\Psi_{C}\rangle=-\frac{4\alpha(Z\alpha)}{3m_h^2}|\Psi_C(0)|^2.
\end{equation}

The Coulomb wave functions $ \Psi_C $  normalized at origin as follows 
\begin{equation}
\label{29}|\Psi_C(0)|^2=\frac{{(Z\alpha)}^3\mu^3}{\pi n^3}\delta_{l0},
\end{equation}
where  $n$ - principal quantum number,  $ \mu$-reduced mass.

Substibuting the norm of  (29) into (28), we get the final form
of the S-level shift of muon atom:

    \begin{equation}
\triangle {E_n}^h=-\frac{4\alpha(Z\alpha)\mu^3}{3n^3m_h^2},
\end{equation}
где $m_h^2\cong 4m_\pi^2$.
 
  Now we calculate numerical values of these corrections for $ n=2 $ (i.e.,
the part of Lamb shift ) in the following concrete cases:

          a) heavy muon atom   $M>>m_\mu$. The reduced mass is equal   $\mu=\frac{Mm_\mu}{M+m_\mu}=m_\mu$  , the shift   is equal  
\begin{equation}
\triangle E^h=-Z^4\frac{\alpha^5m_\mu^3}{6\pi(4m_\pi^2)}\cong -Z^4\cdot 1.67\cdot 10^{-5}\hspace{0,2cm}  {eV}.
\end{equation}

  The total Lamb shift in the muon hydrogen is
   \begin{equation}
   \triangle E_L=(0.2108 \pm 0.0001)\hspace{0,2cm}  {eV}.
  \end{equation}

     b) $(\pi\mu)$-atom .
In this case    $\mu=m_\pi m_\mu/(m_\pi+m_\mu)$  , \hspace{0,3cm}Z=1  and
\begin{equation}
\triangle E^h=\frac{\alpha^5\mu^3}{6\pi(4m_\pi^2)}\cong -0.31\cdot 10^{-5}\hspace{0,2cm} {eV} 
\end{equation}

  For comparison note that the total Lamb shift for   $(\pi\mu)$-atom is equal \cite {17}
\begin{equation}
\triangle E_L= -0.0795 \hspace{0,2cm}   {eV}.
\end{equation}

\section*{5.  CONCLUSION }

\hspace{1cm} It is well known  \cite {18}, that the contribution of the vacuum polarization in the Lamb shift of electronic atoms so small and is about  $\sim1 \%$ of total shift. On the contrary, in the case of muonic atoms the contribution of the vacuum polarization dominates and constitutes $\sim95 \%$  to the total shift. 
Thus it is very important to increase the accuracy of corresponding experimental measurements in order to reveal the corrections by HVP. 
The suggested experiment of measuring Lamb shift on muonic atoms , particularly,  in muonic hudrogen and $(\pi\mu)$-atom \cite {19} , will provide us a very valuable information on the influence of the effects of strong  interactions upon QED at low energies .For example, in $(\pi\mu)$-atom the contribution of the size of pion to the total shift constitutes   $\sim1 \%$ , and the measurement of Lamb shift to  $10^{-3}$ accuracy will lead to the more precise value of pion radius.

\hspace{1cm}The contribution of  HVP has been observed in the anomalous magnetic
moment of muon and in the annihilation cross section\cite {20}   of $ e^+e^-$ into $ \mu^+\mu^-$. Increasing the accuracy of measurements and theoretical calculations of energy lavels of muonic atoms should also reveal the effect of HVP. 

\hspace{1cm} In conclusion, the authors would like to thank  N.N.Bogolyubov, S.B. Gerasimov, A.B. Govorkov, V. A. Meshcheryakov, R. M. Muradyan, L. A. Slepchenko,
A.N. Tavkhelndze for useful discussions. One of   authors / A.K. / expresses gratitude to U.G. Gulyamov, L.Sh. Khodjaev for support and constant interes in work.

The manuscript entered to the publishing department on February 8, 1978 

\begin{thebibliography}{99}
\bibitem{1} Bogolyubov N.N.,Shirkov D.V. Introduction to quantum field theory.  		"Nauka", М., 1976.
\bibitem{2} Faustov R.N. Lectures in School JINR-CERN. preprint JINR , Е2-8786, Dubna, 1975.
            Brodsky S. Proc. of SLAC Summer Inst, on Particle Phys., Stanford, 1973, SLAC- 167, v. 2, p. 141.
\bibitem{3} Auslender  V.L. e.a. ЯФ, 1969, 9, p. 114.
            Benaksas D. е.a. Phys.Lett., 1969, 39В, р.289.
\bibitem{4} Siegrist J. е.a. Phys.Rev.Lett., 1976, 36, p.700.
            Burmester J. e.a. Phys.Lett., 1977, 66B, p.395.
\bibitem{5} Bernardini M. e.a. Phys.Lett., 1974, 51B, p.200.
            Perl M". SLAC-PUB-1614 (1975).
            Schwitters R. SLAC-PUB-1998 (1977). Reviews of Particle Properties. Rev.Mod.Phys., 1976, v.48, no.2.
\bibitem{6}  Gerasimov S.B. . In contribution of seminar on vector mesons and electromagnetic interactions . Dubna, 1969. JINR, Preprint  Р2-4816, p.367. Sakurai J.J.Phys. Lett., 1973, 46В, р.207.
            Greco M. Preprint LNF-74/59 (1974), Frascati.
\bibitem{7} Perl M. e.a. SLAC-PUB-1997 (1977).
\bibitem{8} Renard F.M. Phys.Rep., 1977, V.31C, no. 1.
            Wiik B.H., Wolf G. DESY, 77/01 (1977).
\bibitem{9} Gounaris G.J., Sakurai J.J. Phys.Rev.Lett., 1968, 21, p. 244.
\bibitem{10} Dubnicka S., Meshcheryakov V.A. a) Nucl.Phys., 1974, B83, p.311; JINR, E2-7508, Dubna, 1973. b) Preprint IС/76/102, Trieste, 1976.
\bibitem{11} Matveev V.A., Muradyan R.M., Tavkhelidze A.N. Lett. Nuovo Cim., 1973, 7, p. 719:
             Brodsky S., Fiirrat G. Phys.Rev.Lett., 1973, 31, p.1153
\bibitem{12} Appelquist Т., Georgi H. Phys.Rev., 1973, D8, p.4000.
             Zee A. Phys.Rev., 1973, D8, p.4038.
\bibitem{13} Bailey J. e.a. Phys.Lett., 1977,68В, р.190.
\bibitem{14} Calmet J. e.a. Rev.Mod.Phys., 1976, v.49, no. 2.
\bibitem{15} Gourdin M., de Rafael E. Nucl.Phys., 1969, BIO, p.667.
\bibitem{16} Barger V. e.a. Phys.Lett., 1975, 60B, p.89.
             Startsev С.А. Collection FIAS(FIAN), v.95, "Nauka", М., 1977.
\bibitem{17} Bar-Gadda U., Cho C.F. Phys.Lett., 1973, 46В, р.95.
\bibitem{18} Faustov R.N. EPAN (ЭЧАЯ), Atomizdat, М., 1972, v.3, .1.
\bibitem{19} Schwarts M. XVIII International Conference on High Energy Physivs, Tbilisi, 1976. JINR, Dl,2-10400, Dubna, 1977.
             Scheck F. Acta Physica Austriaca Suppl., 1977, XVIII, p. 629-675.
\bibitem{20} Berends F.A., Komen G.J. Phys.Lett., 1976, 63B, p. 4 32.
\bibitem{21} Chetyrkin K.G., Krasnikov N.V. Nucl.Phys., 1977, B119, p. 174.
\end{thebibliography}
\end{document}